\documentclass[twocolumn,floatfix, showpacs, nofootinbib]{revtex4}

\usepackage[final]{epsfig}
\usepackage{amsmath}
\usepackage{parskip}
\usepackage[dvips]{color}

\definecolor{Joerg}{rgb}{0.0, 0.0, 1.0}

\begin{document}

\title{From In-In to Pb-Pb collisions: Lessons from high-precision Dilepton Measurements}

\author{Thorsten Renk}
\email{trenk@phys.jyu.fi}
\affiliation{Department of Physics, PO Box 35 FIN-40014 University of Jyv\"askyl\"a, Finland}
\affiliation{Helsinki Institute of Physics, PO Box 64 FIN-00014, University of Helsinki, Finland}
\author{J\"org Ruppert}
\email{ruppert@phy.duke.edu}
\affiliation{Department of Physics, Duke University, Box 90305,  
             Durham, NC 27708, USA}

%\author{Berndt M\"uller}
%\email{muller@phy.duke.edu}
%\affiliation{Department of Physics, Duke University, Box 90305,  
%             Durham, NC 27708, USA}

\preprint{DUKE-TH-????}

\begin{abstract}
The NA60 experiment has studied low mass muon pair production in 158 AGeV In-In collisions with unprecedented precision.
With these measurements one can constrain the in-medium modifications of hadrons more 
thoroughly than from previous experiments. We present a model description  which is able to describe the data  well.
We argue that model constraints could be significantly further tightened and theoretical uncertainties in modeling of the medium evolution reduced by an additional experiment measuring  low mass di-muon production in Pb-Pb collisions at 158 AGeV with the same resolution as provided by NA60, thus allowing to disentagle the in-medium spectral function from effects of the evolution model with greater precision.
\end{abstract}

\maketitle

\section{Introduction}
\label{sec_introduction}

The changes of properties of vector mesons (most prominently the $\rho^0$) in a hot and dense medium have been studied extensively in model calculations since their electromagnetic decay into lepton pairs makes them experimentally accessible. Hadronic in-medium properties have also been linked to the chiral phase transition, since vector and axial-vector correlators are predicted to merge as the chirally restored phase is approached. 
However, the ideas about how this is realized in detail diverge.
Brown and Rho \cite{Brown:1991kk,Brown:1995qt,Brown:2001nh,Brown:2002is} proposed that the restoration of chiral symmetry might manifest itself in a scaling of the vector meson masses with the order parameter of the phase transition. On the other hand hadronic many-body calculations differ in the predicted mass shifts but agree on the conclusion that the $\rho$-meson spectral function is significantly broadened due to its interactions with the surrounding medium \cite{Rapp:1999ej,Rapp:1999us,Klingl:1997kf,RSW,Renk:2003hu,RR}. Early QCD sum rule calculations seemed to validate dropping mass scenarios \cite{Hatsuda:1991ez,Asakawa:1993pq}, but QCD sum rules have later been shown to constrain the correlation between in-medium width and mass and not on one of the quantities alone \cite{Leupold:1997dg,BR}. Ultimately, detailed knowledge of all hadronic sources of dileptons is also crucial to address the question if in high energy nucleus-nucleus collisions new non-hadronic contributions to the measured dilepton spectrum arise, such as radiation from deconfined matter.

The CERES and HELIOS-3 collaborations have shown that central nucleus-nucleus 
collisions exhibit a strong enhancement of low mass dileptons compared to 
scaled proton-nucleus and proton-proton collisions \cite{Masera:1995ck,Agakishiev:1995xb,Agakishiev:1998vt,Lenkeit:1999xu}. Although these measurements
demonstrate that indeed considerable in-medium modifications of the vector meson channel 
occur, their precision is not high enough to conclusively discriminate between 
a wide range of different scenarios \cite{Rapp:1999ej,Renk:2003hu,RR}.  Not even the rather straightforward question whether a mass shift of the $\rho$-meson due to the Brown-Rho scaling law \cite{Brown:1991kk,Brown:1995qt,Brown:2001nh,Brown:2002is} is realized in nature or not can
be answered conclusively \cite{Rapp:1999ej} from these results.

The high precision di-muon data of the NA60 experiment now provide the possibility to put strong constraints on our theoretical understanding of in-medium modifications of vector mesons. Most model calculations for in-medium modifications so far have made the simplifying assumption that the $\rho$-meson is at rest. Thus, the only reasonable comparison between calculations using this assumptions and the data employs the NA60 results at low transverse lepton pair momentum. However, since this data set is not yet published we postpone a systematic and extensive theoretical study comparing different in-medium models to a forthcoming publication \cite{NA60waiting}.
We will nevertheless report some of the results of this study \cite{NA60waiting} here. We find that the NA60 data require considerable in-medium broadening of the $\rho$-meson while strong mass shifts are excluded. Moderate mass shifts are not favored by the NA60 data and can be clearly ruled out if not accompanied by strong broadening.
 
In the present paper we follow a different idea and argue that the spectral function favoured by the data in combination with an evolution model based on a scaled version of a well-constrained Pb-Pb evolution model is able to provide a good qualitative and quantitative description of the NA60 data in all momentum bins. We then turn with the same spectral function to the Pb-Pb system where the evolution model is much better constrained by hadronic data \cite{SPS} and show what additional insights beyond what is seen in In-In could be gained if the Pb-Pb system could be measured with the same precision as In-In.

\section{Calculating the dimuon spectrum}

Schematically, the emitted spectrum of dileptons from an evolving thermalized system can be found from the convolution

%\begin{displaymath}
\begin{eqnarray}
\label{E-1}
\frac{d^3N}{dM dp_T d\eta} = \text{evolution} \otimes \frac{dN}{d^4 x d^4q} \otimes \text{acceptance}
\end{eqnarray}
%\end{displaymath}

where the rate is found from the spectral function $R(q, T, \rho_B)$ as

\begin{eqnarray}
\label{E-Rate}
\frac{dN}{d^4 x d^4q}  = \frac{\alpha^2}{12\pi^4} \frac{R(q, T, \rho_B)}{e^{\beta p_\mu u^\mu}- 1}.
\end{eqnarray}

Here, $q$ is the 4-momentum of the emitted muon pair, $T$ is the temperature of the emitting volume element, $\beta=1/T$ and $\rho_B$ its baryon density. The fireball evolution encodes information on radiating volume, $T$, baryon chemical potential $\mu_B$, transverse flow velocity $v_T$,  longitudinal rapidity $\eta$ and chemical non-equilibrium properties such as pion chemical potential $\mu_\pi$.

In the case of NA60, the acceptance is highly non-trivial (and for the comparison with the measured spectrum, a full Monte Carlo simulation of the detector has been carried out by NA60 with the calculation as input). Thus, while a Lorentz-invariant quantity like the invariant mass spectrum $dN/dM$ is in principle independent from the flow profile of the medium, the resulting invariant mass spectrum after acceptance folding is not: The acceptance has $p_T$ dependence and thus the hardening of the momentum spectra due to flow alters the shape of the invariant mass distribution. Thus, for a meaningful comparison of theoretical spectral functions with the data it is mandatory that both the medium evolution and the acceptance are implemented in more than just a schematic way.

\subsection{The evolution model}

A detailed description of the model used to describe the fireball evolution is found in \cite{SPS}. 
The main assumption for the model is that an equilibrated system is formed
a short time $\tau_0$ after the onset of the collision. Furthermore, we assume that this
thermal fireball subsequently expands isentropically until the mean free path of particles exceeds
(at a timescale $\tau_f$) the dimensions of the system and particles 
move without significant interaction to the detector.

For the entropy density at a
given proper time we make the ansatz 
\begin{equation}
s(\tau, \eta_s, r) = N R(r,\tau) \cdot H(\eta_s, \tau)
\end{equation}
with $\tau $ the proper time as measured in a frame co-moving
with a given volume element, $r$ the radius, $\eta_s$ the spacetime rapidity  and $R(r, \tau), H(\eta_s, \tau)$ two functions describing the shape of the distribution
and $N$ a normalization factor.
We use Woods-Saxon distributions with a $\tau$ dependent scale $R_C(\tau)$ ($H_c(\tau)$) and a thickness parameter $d_{ws}$ ($\eta_{ws}$)
to describe the shapes for a given $\tau$.  For a radially non-relativistic 
expansion and constant acceleration we find
$R_c(\tau) = R_0 + \frac{a_\perp}{2} \tau^2$. $H_c(\tau)$ is obtained
by integrating forward in $\tau$ a trajectory originating from the collision center which is characterized
by a rapidity  $\eta_c(\tau) = \eta_0 + a_\eta \tau$
with $\eta_c = \text{atanh } v_z^c$ where $v_z^c$ is  the longitudinal velocity of that 
trajectory. Since the relation between proper time as measured in the co-moving frame 
and lab time is determined by the rapidity at a given time, the resulting integral is
in general non-trivial and solved numerically (see \cite{SPS} for details).
$R_0$
is determined in overlap calculations using Glauber theory. the initial size of the rapidity interval occupied
by the fireball matter. $\eta_0$ is a free parameter and we choose to use the transverse velocity  $v_\perp^f = a_\perp \tau_f$ and
rapidity at decoupling proper time $\eta^f = \eta_0 + a_\eta \tau_f$ as parameters.
Thus, specifying $\eta_0, \eta_f, v_\perp^f$ and $\tau_f$ sets the essential scales of the spacetime
evolution and $d_{\text{ws}}$ and $\eta_{\text{ws}}$ specify the detailed distribution of entropy density.

For transverse flow we assume a linear relation between radius $r$ and
transverse rapidity $\rho = \text{atanh } v_\perp(\tau) =  r/R_c(\tau) \cdot \rho_c(\tau)$
with $\rho_c(\tau) = \text{atanh } a_\perp \tau$.
We allow for the possibility of accelerated longitudinal expansion
which in general implies $\eta \neq \eta_s$ \cite{SPS}.

The model parameters have been adjusted to hadronic transverse mass spectra and HBT correlation measurements in 158 AGeV Pb-Pb and Pb-Au collisions at SPS. The framework was then used to successfully describe photon and dilepton emission and charmonium suppression \cite{SPS}. Since at this point no freeze-out analysis or HBT correlation data are available from NA60 for the In-In system, our strategy is to start with the parameter set derived for Pb-Pb collisions and use geometrical scaling arguments to go to In-In. 

We infer the change in total entropy production from the ratio of charged particle rapidity densities
$dN_{ch}/d\eta$ at 30\% peripheral Pb-Au collisions with $2.1 < \eta <2.55$ measured by CERES \cite{Agakishiev:1995xb,Agakishiev:1998vt,Lenkeit:1999xu} and semi-central In-In collisions at $\eta=3.8$ measured by NA60 \cite{NA60data}, $dN_{ch}^{In-In}/dN_{ch}^{Pb-Pb} = 0.68$.
The number of participant baryons and the effective initial radius (for the sake of simplicity, we map the overlap area in non-central collisions to a circle with the same area) are obtained with nuclear overlap calculations. We assume that the stopping power (which determines the width of the inital distribution of entropy and baryon number in rapidity) scales approximately with the number of binary collision per participant, $N_{bin}/N_{part}$. This determines $\eta_0$. The shape parameters $d_{ws}$ and $\eta_{ws}$ have no great impact on electromagnetic emission into the midrapidity slice, they primarily govern the ratio of surface to volume emission of hadrons. Thus, we leave them unchanged from their value determined in \cite{SPS}.
Under the assumption that the
physics leading to equilibration is primarily a function of incident energy we keep the formation time $\tau = 1$ fm/c as in
Pb-Pb collisions. We stress that the final results exhibit no great sensitivity to the choice of either
$\eta_0$ or $\tau_0$ except in the high $M$, high $p_T$ limit.

The biggest uncertaintly is the choice of the decoupling temperature $T_F$. Due to the fact that the In-In system is smaller than Pb-Pb, we expect a higher decoupling temperature. However, without simultaneous knowledge of HBT correlations and transverse mass spectra an unambiguous answer cannot be obtained. We tentatively choose $T_F = 130$ MeV. Our lack of detailed knowledge translates primarily into an uncertainty in the normalization of the final results (which depend on the emission duration which increases for lower $T_F$). It is for this reason that the result is normalized to the data in the interval $M<0.9$ GeV in the comparison to account for discrepancies of the order of a few percent.

We fix the flow velocity at $T_C$ by choosing the value of transverse flow obtained at $T = 130$ MeV in our Pb-Pb model in the same centrality class. There is likewise an uncertainty (which is to some degree correlated with the choice of $T_F$) since flow determines the slope of hadronic (and to a lesser degree electromagnetic) $p_T$ spectra.

The equation of state in the hadronic phase and the off-equilibrium parameters $\mu_\pi, \mu_K$ are inferred from statistical model calculations as described in \cite{SPS, Hadronization}. The resulting fireball evolution is characterized by a peak temperature of about $250$ MeV, a lifetime of $\sim 7.5$ fm/c and a top transverse flow velocity of $0.5 c$ at decoupling.

\subsection{The spectral function}
\label{hadronicspec}

Below the phase transition temperature the effective degrees of freedom
are hadrons. The photon couples now to the lowest-lying dipole excitations of the vacuum,
the $\rho$, $\omega$ and $\phi$ mesons and multi-pionic states with the appropriate quantum number.  

There are different effective hadronic models and techniques for the calculation of the properties of hadronic matter near and below the phase boundary. For 
reviews see e.g. \cite{Rapp:1999ej, Alam:1999sc, Gale:2003iz}.
Most of the effective model calculations that are constrained by phenomenological data such as the hadronic and electromagnetic decay widths of the particles predict substantial broadening in matter with comparably small shifts of the in-medium masses.

As already discussed in the introduction  we focus here on one specific hadronic model calculation which is favoured by the data and discuss it as a 'generic' example.
The model calculation employed here solves truncated Schwinder-Dyson equations incorporating a self-consistent resummation of the $\pi-\rho$ interaction in order to determine the thermal broadening of the $\rho$-meson. For a detailed discussion of this approach see \cite{RR} and citations therein \cite{DS1}. This $\Phi$-derivable approach self-consistently takes into account the finite in-medium damping width of the pion in the thermal heat bath which contributes substantially to the broadening of the $\rho$-meson. Effects of baryons are not included in this approximation scheme. We note that the finding that only thermal effects would be able to account for the data is clearly interesting.

As long as the thermodynamically active degrees of freedom are quarks and
gluons, the timelike photon couples to the continuum of thermally
excited $q\overline{q}$ states and subsequently converts into a charged lepton pair.
The calculation of the photon spectral function  at the one-loop level is performed using
standard thermal field theory methods. The leading-order result
for bare quarks and gluons as degrees of freedom is well known \cite{Rapp:1999ej}.
This result, however, holds only
up to perturbative higher order corrections in $g_s$ that take into account
collective plasma effects. We describe those by means of a quasiparticle model, additional details can be found in \cite{RSW}.

\subsection{The vacuum $\rho$-meson}

\label{S-Vacuum-Rho}

In addition to the dilepton radiation from the medium (mainly through the $\rho^0$ channel) there is also radiation from e.m. decays of $\rho$ mesons in vacuum after the thermalized evolution has ceased. In principle there is also radiation from $\omega$ and $\phi$ vaccum decays, however due to the small width of these states this contribution is experimentally very well identified and has been subtracted \cite{NA60data}. The vacuum decay of $\rho$ mesons is commonly referred to as 'cocktail $\rho$' and obtained by calculating the produced number of $\rho$ mesons by a statistical model calculation.

In such a calculation, the number of $\rho$ mesons is found under the assumption that the relative abundances of hadron species are given by the thermal expectation value above a temperature $T_C$ (close to the transition temperature $T_F$) but that subsequently only resonance decays  occur. In such a framework, the decay of heavy resonances leads to an overpopulation of pion phase space as compared to the  pion phase space density in chemical equilibrium. This can be parametrized by the introduction of a pion chemical potential $\mu_\pi$ (which grows with decreasing temperature) and leads via the tightly coupled $\rho \rightarrow \pi\pi \rho$ channel to an effective $\mu_\rho = 2 \mu_\pi$. 

However, while tracking of the resonance decays fixes the number of $\rho$-mesons at thermal decoupling, their momentum spectrum remains unknown. In particular, $\rho$ mesons after chemical decoupling also participate in the collective expansion of matter. While this effect does not influence the integrated invariant mass spectrum, the acceptance leads to a significant influence of flow on the detected signal. Thus, we go beyond the statistical model formulation and compute the full momentum spectrum of $\rho$ mesons after decoupling using the Cooper-Frye formula

\begin{equation}
E \frac{d^3N_{\rho^0}}{d^3p} =\frac{1}{(2\pi)^3} \int d\sigma_\mu p^\mu
\exp\left[\frac{p^\mu u_\mu - \mu_\rho}{T_f}\right]
\end{equation}

with $p^\mu$ the momentum of the emitted $\rho$ and $d\sigma_\mu$ an element of the freeze-out hypersurface (determined by the condition $T=T_F$) just as we calculate the emission of any other hadron from the medium. We let this ensemble undergo electromagnetic decay with vacuum width and lifetime. 

\section{Results}

\begin{figure*}[!htb]
\epsfig{file=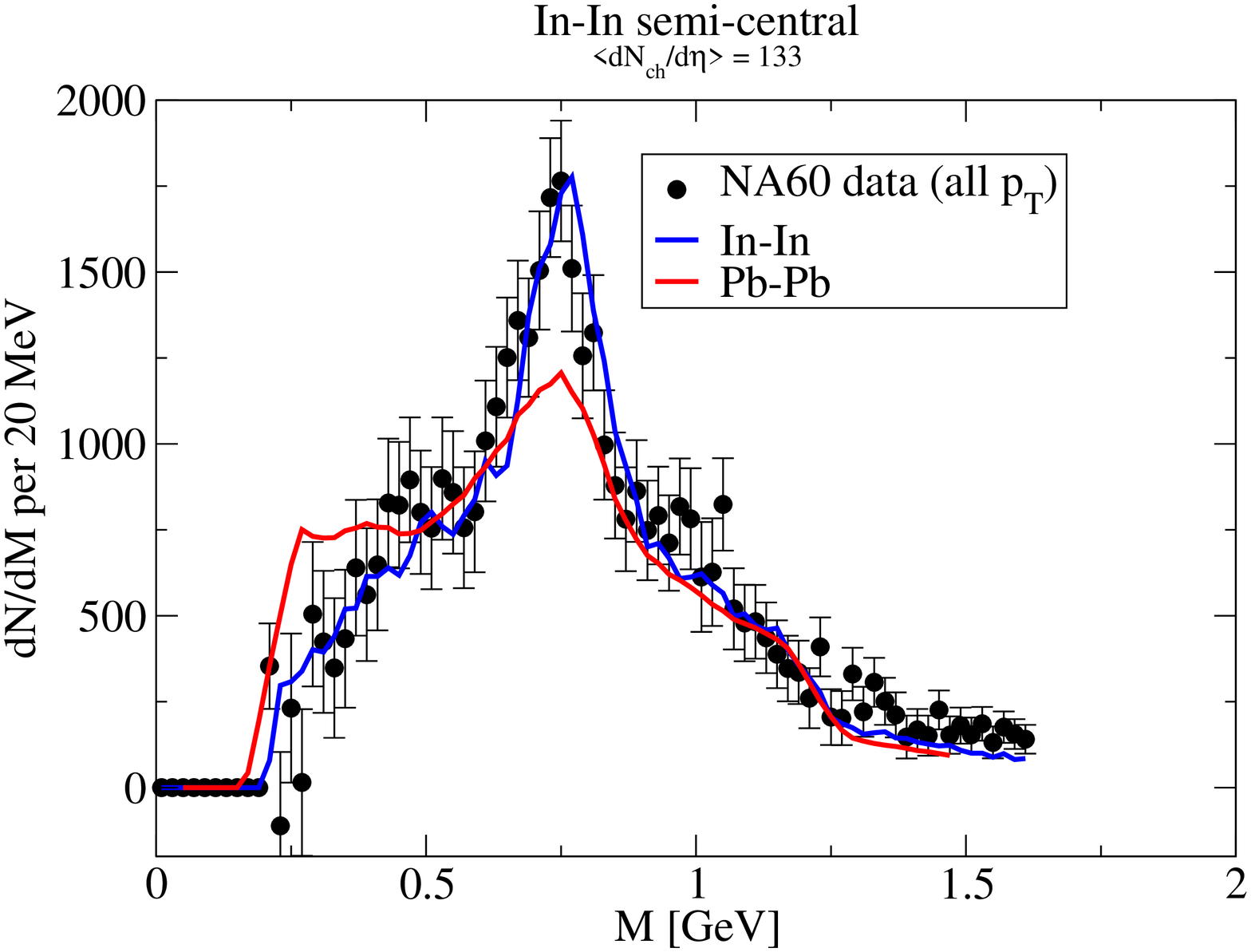, width=8cm}\epsfig{file=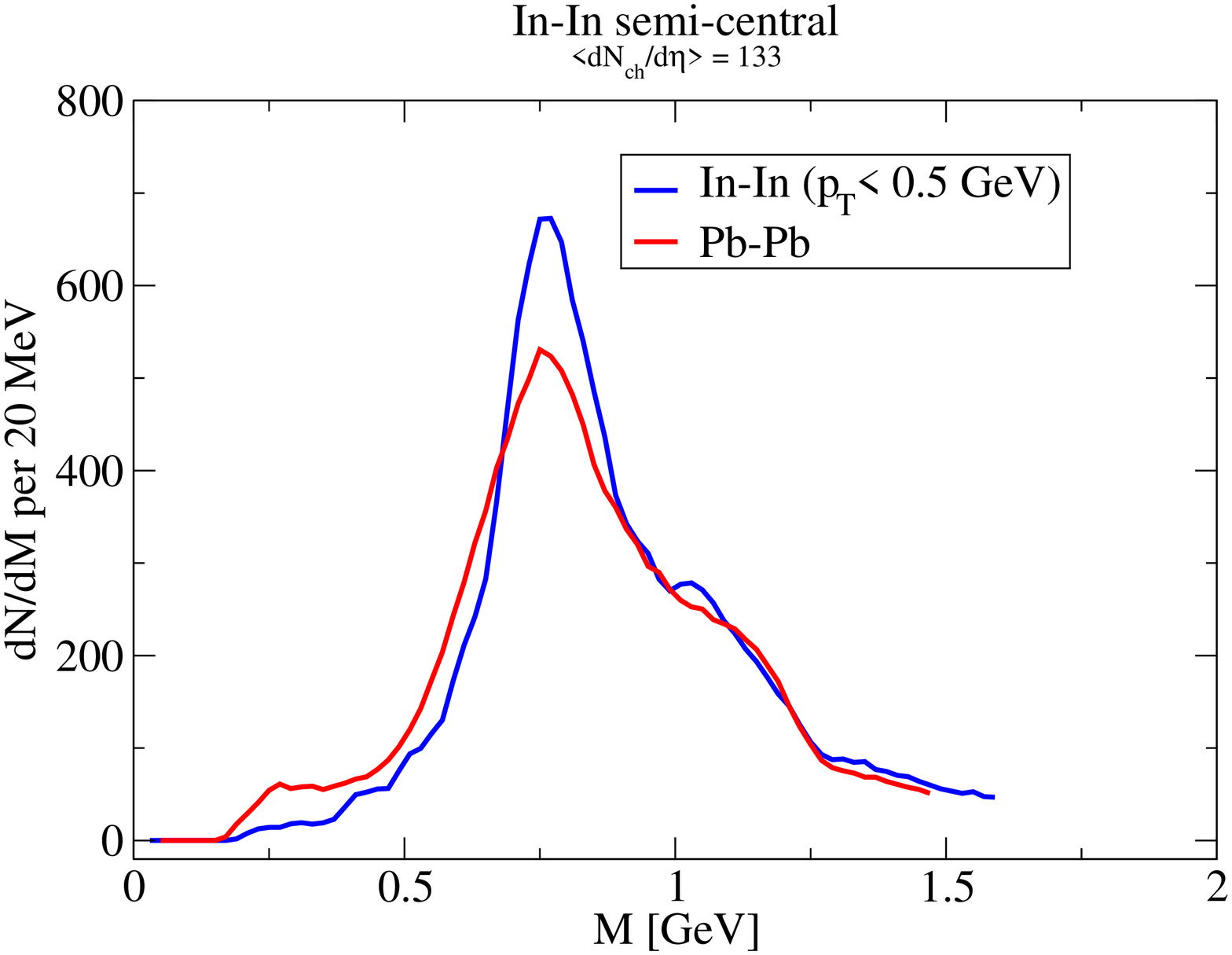, width=8cm}
\caption{\label{F-RR}
Calculations of the di-muon invariant mass spectrum as expected in Pb-Pb and In-In collisions at $158~{\rm AGeV}$ assuming the resolution and accetance of the NA60 detector as obtained
within a self-consistent $\Phi$-functional approach for the $\pi-\rho$ interaction \cite{RR}) compared with the NA60 dimuon measurementsof the In-In system.  For further explanation see main text. 
Left panel: all $p_T$ Right panel: $p_T < 0.5$ GeV. All model calculations have been normalized to the data in the mass interval $M < 0.9$ GeV.}
\end{figure*}

In Fig. \ref{F-RR}, left panel, we present the comparison of the calculation to the NA60 all $p_T$ data for In-In collisions at $158 ~{\rm AGeV}$ using the spectral function obtained in the $\Phi$-functional approach \cite{RR}. The calculation contains the vacuum $\rho$ 
as described in section \ref{S-Vacuum-Rho} and a $D\overline{D}$ background \cite{NA60data}. 
We note that the agreement with the data indicates that the relative contribution of the vacuum $\rho$ and in-medium $\rho$ is at least approximately verified. While the strength of the vacuum $\rho$ parametrically scales like the freeze-out volume times the $\rho$ lifetime $V(\tau_f) \tau^\rho_{vac}$, the in-medium radiation scales like $\int V(\tau) d\tau$ and is thus sensitive to the lifetime of the fireball. Thus, the relative strength is a consistency test for the medium lifetime in the In-In collision evolution model. 

Fig. \ref{F-RR} also presents the calculation of the di-muon spectrum that would be expected in $158 ~{\rm AGeV}$ Pb-Pb collisions with the same detector and resolution as used by the NA60 collaboration in In-In collisions. In order to compare In-In collisions and Pb-Pb collisions we normalized both calculation to the In-In collision data in the mass region below $0.9$ GeV. We note that the model describes the absolute normalization of the In-In data set with less than 5\% discrepancy.

In Fig. \ref{F-RR}, right panel, we compare the predictions for the Pb-Pb system and In-In system for low $p_T$ normalized to NA60 In-In low $p_T$-data (the yet unpublished low $p_T$ data are not shown)\cite{DS2}. This shows that the low $p_T$ cut acceptance of the detector puts serious limitations to the capability of low $p_T$ data to distinguish different lineshapes, in particular in the mass range below the $\rho$ mass.

Comparing both calculations one realizes the following most prominent distinctions between the spectra for Pb-Pb collisions and In-In collisions:
\begin{enumerate}
\item The relative contribution of the in-medium $\rho$ to the vacuum $\rho$ is more pronounced due to the longer lifetime in Pb-Pb. 
\item Contributions at low invariant mass are more emphasized in the Pb-Pb system due to lower $T_F$.  This is clearly seen in the 
all $p_T$ binning but somewhat obscured for the low $p_T$ data.
\end{enumerate} 

These features are a clear test for different evolution models and in particular for differences in lifetime and $T_F$. They show that the in-medium modifications
of the $\rho$-meson would show up more prominently in a $158~{\rm AGeV}$ Pb-Pb measurement with the same resolution than in In-In,
since the relative contribution of the vacuum $\rho$-meson would be reduced. 
The Pb-Pb collision experiment would therefore probe the in-medium modifications in the mass region around and below the vacuum rho-mass 
more clearly than in the In-In system at all and high $p_T$. However, it would also probe these spectral functions at on average lower temperature. This could provide information of 
how the the in-medium properties vary in the region between the different $T_F$ of the Pb-Pb and In-In systems, going beyond what can be constrained by In-In alone. 

A third distinction, though expected, is not visible in the results. Since the radiation from hot volume elements with $T>T_C$ is increasingly important for the contribution to invariant masses above the vacuum $\rho$-mass, it is therefore  expected that the higher initial temperature of the Pb-Pb system leads to a {\it harder} slope of the Pb-Pb di-muon spectrum at high invariant masses where more relative contribution to the total yield is expected
from above $T>T_C$ than in In-In.  Although this is true when comparing the spectra before acceptance, after acceptance correction the effect is not significant in the all $p_T$ data.

\section{Conclusions and outlook}

In this paper we studied the di-muon production in $158~{\rm AGeV}$ In-In collisions measured by the NA60 collaboration. We first presented a model which reproduces the In-In data qualitatively and quantitatively well. We then argued that a measurement of Pb-Pb with the same accuracy would be capable of providing additional non-trivial information and so help to disentangle effects of spectral function and evolution model. Already the $p_T$-integrated lineshape indicated sizeable differences which in the simulation arise from different lifetimes and freeze-out temperatures. It is of considerable advantage that the dynamics of Pb-Pb collisions is already well constrained from measurements of hadronic observables. 
The expectation in the high mass region would also differ in comparison between Pb-Pb and In-In collisions. We argued that this could provide
important information about contributions from non-hadronic sources. This is especially important since a purely hadronic evolution seems unable to reproduce the NA60 In-In data in the high $p_T$ high $M$ region \cite{NA60waiting}  which is an unexpected and interesting hint that even in the rather small In-In system some amount of partonic evolution may contribute to the observed radiation. This contribution should be increased in Pb-Pb collisions and be visible in the Pb-Pb to In-In comparison if a high $p_T$ binning is performed.

We will present a detailed systematic and extensive theoretical study comparing different in-medium models and their predictions for different $p_T$ cuts for the In-In system in the near future \cite{NA60waiting}.

\begin{acknowledgments}
We are grateful to Berndt M\"uller for many valuable comments and discussions and his support of this work.
We thank Sanja Damjanovic for taking care of the NA60 acceptance simulations with our model input and (together with Hans Specht) for many interesting discussions and Carlos Lourenco for comments and discussions.
This work was supported by DOE grant DE-FG02-05ER41367, the Academy of Finland, Project 206024 and the Alexander von Humboldt Foundation's Feodor Lynen Fellow program.
\end{acknowledgments}

\bibliography{u4}

\begin{thebibliography}{27}
\bibitem[{DS2()}]{DS2}
\bibinfo{note}{We mention that the results of our calculation for the
  $\Phi$-derivable approach's hadronic spectral function as input for the
  evolution model presented here is in very good agreement not only with the
  all $p_T$ NA60 data \cite{NA60data}, but also with the low $p_T$ and high
  $p_T$ cuts of the preliminary data \cite{Sanja2}. Those are not yet publicly
  available. A detailed comparison with those will therefore been postponed to
  the extensive systematic study mentioned in the introduction
  \cite{NA60waiting}.}
\end{thebibliography}

\end{document}